# The performance analysis of a Quantum-Mechanical Carnot-like Engine using Diatomic Molecules.

E. O. Oladimeji[†1,2]; T. T. Ibrahim[1]; A.N. Ikot[3]; J.D. Koffa[1]; V. T. Idundun[1]; E. C. Umeh[1]; J.O. Audu[1,4]

1. *Theoretical Physics Group, Department of Physics, Federal University Lokoja, Lokoja, Nigeria.*
2. *Institute of Physical Research and Technology, Peoples' Friendship University of Russia, Moscow, Russia.*
3. *Theoretical Physics Group, Department of Physics, University of Port Harcourt, Port Harcourt, Nigeria.*
4. *ICTP East African Institute for Fundamental Research (EAIFR), University of Rwanda, Kigali, Rwanda.*

**ABSTRACT**

This study presents an analysis of a quantum-mechanical formulation of the Carnot-like cycle using diatomic molecules- the Morse Oscillator, as the working substance. The generalized model with an arbitrary one-dimensional potential is used to obtain the important performance parameters such as the efficiency, the power output, and the optimal region of the engine by considering well width $L$ moving with finite speed. The optimal efficiency, the maximum power output, and dimensionless power ranges of the working substance was also determined. The results obtained in this work are found to agree with those obtained for similar engine but with different working substances.

---

*† Corresponding author: E. O. Oladimeji. e-mail:* [nockjnr@gmail.com](nockjnr@gmail.com)

# I. INTRODUCTION

It is well known that classical heat engines [CHEs] are designed to convert heat energy to useful mechanical work by subjecting the working substance through a cyclic process between two reservoirs at different temperatures. An ideal heat engine is expected to be highly efficient and to possess large power output while it exhibits small power fluctuations. However, in reality [CHEs] have obvious limitations. Their efficiency and of course overall performance are reduced by irreversible losses since the total energy derived from the reservoir at a higher temperature is not completely converted into mechanical work. This fact is evident in the low-efficiency of the CHEs noted in Refs [1–4]. In the general case, the expression of the efficiency is defined as:

$$\eta = 1 - \frac{E_L}{E_H}$$

where $E_L$ ($E_H$) is the expectation value of the system Hamiltonian along the analogue of the isothermal process at high (low) "temperature."

Today, the miniaturization of these heat engines from macroscale to nanoscale where quantum fluctuation takes charge of the system has made the limitation of classical heat engines at this state inconsequential [5,6]. Recent successful experiments of Nanoscopic heat engine [7–13], have begun to justify several theoretical efforts [14–19] that have been formulated since the discovery of the relation between the efficiency of the Carnot engine and the 3-level maser by *Scovil and Schulz-DuBois* in 1959 which can be referred to as the birth of quantum heat engines [QHEs] [20,21]. These theoretical studies have led to the use of several systems to mimic the working substances of the QHEs. Examples of such systems include the free particle [FP] in the box proposed by *Bender et al.* [22–26] and others such as spin systems [27–29], two-level or multilevel systems [19,25], cavity quantum electrodynamics systems [30,31], coupled two-level systems [51], Harmonic oscillators[HO] [32–35], Pöschl-Teller Oscillator[PTO] [36–39], Wood-Saxon [23], Shortcut-to-Adiabaticity [40,41], single trapped ion [2], optomechanical systems [42–45], quantum dots [46], cold bosons [47] etc.

In the [FP] model of *Bender et. al.,* [25,26] a single particle of mass m in an infinite one-dimensional potential well of width $L$ is taken as the working substance which undergoes a Carnot cycle. The walls of the confining potential and energy (i.e., the pure-state expectation value of the Hamiltonian) are also taken to represent the functions of the piston and the temperature as in classical thermodynamics, respectively. Following this formulation, a Quantum Carnot-like heat engine whose working substance couples to the energy bath in an isoenergetic process was investigated by *Abe* [48,49]. In the isoenergetic process, the temperature is replaced with heat fluxes which produces work by changing the potential width $L$ at a slow speed. Therefore, instead of the temperature variable in classical thermodynamics, the quantum mechanical expectation value of the Hamiltonian is used during the heat cycles of the [QHE].

Optimization of the isoenergetic process reveals the possible finite nature of the movement of the well width [48], in analogy to the piston's finite speed achievable via finite time thermodynamics. This was furthered by *Wang et al* [22] who analysed similar process for two particles and three levels system resulting in an improved power output considering the potential for energy leakage $Q_r$ between the two energy baths. The generalization of this problem was explained in [50], when the performance of a multilevel quantum heat engine of an ideal N-particle Fermi system was observed [51]. Although, other literature has shown reservation on the isoenergetic process due to its violation of the law of equipartition of energy in the quantum regime [52]. However, quantum mechanics is known to explore novel concepts that have no exact analogue in classical thermodynamics i.e., the dephasing bath and energy bath [53], but it is still necessary to explicate the implications of isoenergetic process to thermodynamics [54].

In this work, we introduce the Morse oscillator potential [MP] as a working substance to construct an adiabatic and isoenergetic quantum analogous process to analyse the efficiency and performance analysis of an idealized reversible heat engine i.e., Carnot engine, following a pedagogical approach similar to *Bender et al* which has been a very useful tool with many researchers [48,49,55–59]. This potential is one of the simplest and most "realistic" anharmonic potential models. Since its introduction by *P.M Morse* in 1929, it has been applied to various problems in several fields of Physics, particularly in model calculations in spectroscopy, diatomic molecule vibrations and scattering etc. The potential is [60]:

$$V(x) = D_0 \gamma (\gamma - 2)$$

given that $\gamma = e^{-\alpha(x-x_0)}$, where $D_0$ is the depth of the potential energy well, $x_0$ is the equilibrium internuclear separation, $\alpha$ is a positive characteristic parameter of this potential and $x$ is the radial part of the spherical coordinate (in 1-dimension). This potential has the minimum, $-D_0$, at $x = x_0$. The potential width, $L$, at the value $V = -V_0$ ($0 < V_0 < D_0$) is $L = (1/\alpha) \ln(\rho_+/\rho_-)$ where $\rho_\pm \equiv 1 \pm \sqrt{1 - V_0/D_0}$ and the energy eigenvalues of the bound states are given as follows [61–64]:

$$E_n^{MP}(L) = \hbar\omega\left(n + \frac{1}{2}\right) - \frac{\left[\hbar\omega\left(n + \frac{1}{2}\right)\right]^2}{4D_0} \quad (1)$$

where $n = 0,1,2,3,...$ and $\omega = s\pi c/L$.

The formal relation between pressure $\hat{P}$ and its energy $\hat{E}$ operator or the Hamiltonian is $\hat{P}(\hat{x},\hat{p},L) = -(\partial/\partial L)H(\hat{x},\hat{p},L)$, where the relation between the pressure $\hat{P}$ and energy $\hat{E}$ operators is defined as:

$$\hat{P}_n(L) = -\frac{\partial \hat{E}_n}{\partial L}$$

Therefore, its Pressure operator $\hat{P}$ becomes:

$$P_n^{MP}(L) = \frac{A}{L^2}\left(n + \frac{1}{2}\right) - \frac{\left[A\left(n + \frac{1}{2}\right)\right]^2}{2D_0 L^3} \quad (2)$$

where $n = 0,1,2,3...$ and $A = \hbar s\pi c$.

This paper is organized as follows: In sec. II, a model of a quantum Carnot-like engine consisting of two adiabatic and two isoenergetic processes using Morse oscillator potential [MP] as the working medium is established, and its efficiency $\eta$ is derived. The quantum characteristic of the working medium is studied, and the optimal relationship between the dimensionless power output $p^*$ versus the efficiency $\eta$ for the quantum Carnot engine was analysed and discussed in sec. III. In sec. IV, we reduced the derived efficiency to that of the Harmonic Oscillator [HO], compared our derived result with other known models and presented our conclusion in sec. V.

## II. THE CARNOT-LIKE CYCLE

The Carnot cycle is the most ideal cycle model in engine research which consists of four branches, two of which are adiabatic and two isothermal strokes. This cycle serves as a template for all reversible engines. To achieve the quantum description of this cycle, its cylinder is replaced by a potential well likewise, the fluid (an ideal gas) and temperature $T$ are replaced with the Morse oscillator potential [MP] and the expectation value of the Hamiltonian $E$, respectively. However, our Quantum heat engine does not possess Temperature $T$ and heat baths but rather heat fluxes and energy baths, respectively. This implies that our engine commences its procedure rather with an isoenergetic expansion.

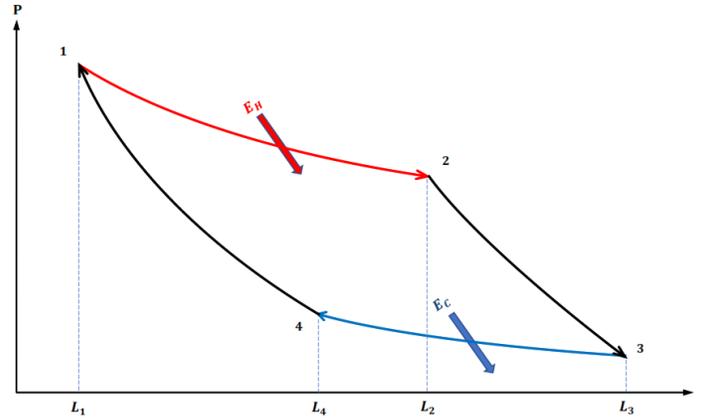

*Figure 1: The schematic representation of a quantum heat engine cycle in the plane of the width (L) and Pressure P(L). The cycle consists of two (2) isoenergetic and two (2) adiabatic processes.*

During the expansion, the quantum system at the initial state $\psi(x)$ of volume $L$ is a linear combination of eigenstates $\phi_n(x)$, with the expectation value of the Hamiltonian remaining constant as the walls of the well move. The expansion coefficient $a_n$ changes such that $E(L)$ remain fixed as $L$ changes:

$$E(L) = \sum_{n=0}^{\infty} |a_n|^2 E_n$$

where $E_n$ is our diatomic energy spectrum and the coefficients $|a_n|^2$ are constrained by the normalization condition $\sum_{n=0}^{\infty} |a_n|^2 = 1$.

### 2.1 Isoenergetic Expansion

In the first stage, the system experiences an isoenergetic expansion from its initial state $n = 0$ at point 1 (i.e. from $L = L_1$ to $L = L_2$) and into the second state $n = 1$, Thus, the state of the system is a linear combination of its lowest-two possible energy eigenstates $\Psi_n$, while its expectation value of the Hamiltonian remains constant.

At $n = 0$,

$$E_H(L) = \frac{A}{2L_1} - \frac{A^2}{16L_1^2 D_0} \tag{3}$$

where $E_H$ denotes the expectation value of the system Hamiltonian along the isoenergetic process at high energy, while the eigenstates:

$$\Psi_n = a_0(L)\phi_0(x) + a_1(L)\phi_1(x)$$

where $\phi_0$ and $\phi_1$ are the wavefunctions of the ground and first excited states, respectively.

$$E(L) = \sum_{n=0}^{1}(|a_0|^2 + |a_1|^2)E_n = |a_0|^2 E_0 + |a_1|^2 E_1$$

$$E(L) = \left(\frac{A}{2L} - \frac{A^2}{16L^2 D_0}\right)|a_0|^2 + \left(\frac{3A}{2L} - \frac{A^2}{16L^2 D_0}\right)|a_1|^2$$

Using the boundary condition $|a_0|^2 + |a_1|^2 = 1$, The expectation value of the Hamiltonian in this state as a function of $L$ is given as $E = \langle\psi|H|\psi\rangle$:

$$E(L) = \frac{A}{2L}(3 - 2|a_0|^2) - \frac{A^2}{16D_0 L^2}(9 - 8|a_0|^2)$$

Setting the expectation value to be equal to $E_H$ i.e., when $n = 0$

$$\frac{A}{2L_1} - \frac{A^2}{16L_1^2 D_0} = \frac{A}{2L}(3 - 2|a_0|^2) - \frac{A^2}{16D_0 L^2}(9 - 8|a_0|^2)$$

Therefore, by comparison, the maximum value of $L = L_2$ when $|a_0|^2 = 0$:

$$L_2 = 3L_1 \tag{4}$$

The pressure during the isoenergetic expansion is:

$$P_1(L) = \frac{A}{2L_1 L} - \frac{A^2}{8L_1^2 L D_0} \tag{5}$$

The product $LP_1(L) = constant$. This is an exact quantum analogue of a classical *equation of state*.

### 2.2 Adiabatic Expansion

Next, the system expands adiabatically from $L = L_2$ until $L = L_3$. During this expansion, the system remains in the first excited state $n = 1$ as no external energy comes into the system and the change in the internal energy equals the work performed by the walls of the well. The expectation value of the Hamiltonian is:

$$E(L) = \frac{3A}{2L} - \frac{9A^2}{16L^2 D_0} \tag{6}$$

The pressure $P$ as a function of $L$ is:

$$P_2(L) = \frac{3A}{2L^2} - \frac{9A^2}{8L^3 D_0} \tag{7}$$

The product $L^2 P_2(L)$ in (7) is a constant and it is considered as the quantum analogue of the classical *adiabatic process*.

### 2.3 Isoenergetic Compression

The system is in the second state $n = 1$ at point 3 and it compresses Isoenergetically to the initial (ground) state $n = 0$ (i.e., from $L = L_3$ until $L = L_4$) as the expectation value of the Hamiltonian remains constant. Thus, the state of the system is a linear combination of its two energy eigenstates.

At $n = 1$,

$$E_L(L) = \frac{3A}{2L_3} - \frac{9A^2}{16L_3^2 D_0} \tag{8}$$

where $E_L$ denotes the expectation value of the system Hamiltonian along the isoenergetic process at low energy, while the eigenstates:

$$\Psi_n = b_0(L)\phi_0(x) + b_1(L)\phi_1(x)$$

where $\phi_0$ and $\phi_1$ are the wave functions of the ground and first states respectively

$$E(L) = \sum_{n=0}^{\infty}(|b_0|^2 + |b_1|^2)E_n = |b_0|^2 E_0 + |b_1|^2 E_1$$

$$E(L) = \left(\frac{A}{2L} - \frac{A^2}{16L^2 D_0}\right)|b_0|^2 + \left(\frac{3A}{2L} - \frac{A^2}{16L^2 D_0}\right)|b_1|^2$$

The coefficients satisfy the condition $|b_0|^2 + |b_1|^2 = 1$. The expectation value of the Hamiltonian in this state as a function of $L$ is calculated using $E = \langle\psi|H|\psi\rangle$, which results in:

$$E(L) = \frac{A}{2L}(1 + 2|b_1|^2) - \frac{A^2}{16L^2 D_0}(1 + 8|b_1|^2)$$

Setting the expectation value to be equal to $E_L$ i.e., $n = 1$

$$\frac{3A}{2L_3} - \frac{9A^2}{16L_3^2 D_0} = \frac{A}{2L}(1 + 2|b_1|^2) - \frac{A^2}{16L^2 D_0}(1 + 8|b_1|^2)$$

Therefore, by comparison, the maximum value of $L = L_4$ when $|b_1|^2 = 0$:

$$L_4 = \frac{1}{3} L_3 \tag{9}$$

The pressure during the isoenergetic compression is:

$$P_3(L) = \frac{3A}{2L_3 L} - \frac{9A^2}{8D_0 L_3^2 L} \tag{10}$$

The product $LP_3(L) = constant$. This is an exact quantum analogue of a classical *equation of state*.

### 2.4   Adiabatic Compression

Finally, the system returns to the initial state $n = 0$ at point 4 as it compresses adiabatically (i.e., from $L = L_4$ until $L = L_1$). The expectation of the Hamiltonian is given by:

$$E(L) = \frac{A}{2L} - \frac{A^2}{16L^2 D_0} \tag{11}$$

and the pressure applied to the potential well's wall $P$ as a function of $L$ is:

$$P_4(L) = \frac{A}{2L^2} - \frac{A^2}{8L^3 D} \tag{12}$$

The product $L^2 P_4(L)$ in (12) is a constant and it is considered as the quantum analogue of the classical *adiabatic process*.

The work $W$ performed by the quantum heat engine during one closed cycle, along the four processes described above is the area of the closed loops represented in *Fig.1*. By using equ. (5), (7), (10) and (12) one obtains

$$W = W_{12} + W_{23} + W_{34} + W_{41}$$

$$W = \int_{L_1}^{L_2} P_1 dL + \int_{L_2}^{L_3} P_2(L) dL + \int_{L_3}^{L_4} P_3 dL + \int_{L_4}^{L_1} P_4(L) dL$$

$$W = \frac{A}{2} \left\{ \left( \frac{L_3 - 3L_1}{L_1 L_3} \right) + \frac{A}{4D_0} \left( \frac{9L_1^3 - L_3^3}{L_1^2 L_3^2} \right) \right\} ln3 \tag{13}$$

the heat quantity $Q_H$ absorbed by the potential well during the isoenergetic expansion in a quantum engine:

$$Q_H = \left( \frac{A}{2L_1} - \frac{A^2}{8L_1^2 D_0} \right) \tag{14}$$

The efficiency of the heat engine is defined to be $\eta = \frac{W}{Q_H}$, therefore.

$$\eta = 1 - \frac{\left(\frac{3A}{2L_3} - \frac{9A^2}{8L_3^2 D_0}\right)}{\left(\frac{A}{2L_1} - \frac{A^2}{8L_1^2 D_0}\right)} \tag{15}$$

$$\eta = 1 - \left(\frac{3L_1}{L_3}\right)\frac{(1-\alpha)}{(1-\beta)}$$

where $\alpha = \frac{3A}{4L_3 D_0}$ and $\beta = \frac{A}{4L_1 D_0}$.

Considering the HO-limit of the Morse potential [62], where the depth of the potential energy well $D_0 \to \infty$, this implies that $\alpha$ and $\beta$ become zero and equation (15) becomes:

$$\eta = 1 - \left(\frac{3L_1}{L_3}\right) \tag{16}$$

Given that $r = L_3/L_1$, where $L_1$ and $L_3$ are assumed to be fixed and variable parameters respectively, this implies that the efficiency is:

$$\eta = 1 - \frac{3}{r}$$

where:

$$r > 3 \tag{17}$$

The derived efficiency for this quantum heat engine is synonymous with that of the efficiency for the heat engine working with a one-dimensional harmonic oscillator potential. Substituting the equ. (3) and (8) into (15), the efficiency can be written as:

$$\eta = 1 - \frac{E_L}{E_H} \tag{18}$$

Note that this efficiency is analogous to that of a classical thermodynamic Carnot cycle.

### III.   OPTIMIZATION OF THE PERFORMANCE OF THE HEAT ENGINE

We will now discuss the quantum heat engine's power output. The introduction of power indicates the inclusion of time into the system i.e., the system's cycle is finite, which in turn denies the introduction of heat fluxes due to the absence of a heat bath [22]. The introduced time scale associated with the variation of the state is taken to be larger than that of the dynamical one $\sim \hbar/E$ to comply with the adiabatic theorem. The potential wall moves at a small but infinite speed, while $\bar{v}(t)$ and $\tau$ is defined as the average speed of change of $L$ and the total cycle time, respectively. The total amount of movement during a single cycle, $L_0$, is given by:

$$L_0 = (L_2 - L_1) + (L_3 - L_2) + (L_3 - L_4) + (L_4 - L_1) = 2(L_3 - L_1)$$

The speed $\bar{v}(t)$ is expected to be slow enough so that the variation of $L$ is much slower compared with the dynamics time scale $\sim \hbar/E$. The total cycle time $\tau$ can be given as:

$$\tau = \frac{L_0}{\bar{v}} = \frac{2(L_3 - L_1)}{\bar{v}} \tag{19}$$

Thus, the power output $p$, after a single cycle, is:

$$p = \frac{W}{\tau} = \frac{A\bar{v}}{4}\left\{\left(\frac{L_3}{L_1} - 3\right)\frac{1}{L_1^2\left(\frac{L_3^2}{L_1^2} - \frac{L_3 L_1}{L_1^2}\right)} + \frac{A}{4D}\left(9 - \frac{L_3^2}{L_1^2}\right)\frac{1}{L_1^3\left(\frac{L_3^3}{L_1^3} - \frac{L_3^2 L_1}{L_1^2}\right)}\right\} ln3$$

$$p = \left\{\frac{A\bar{v}}{4L_1^2}\frac{(r-3)}{(r^2-r)} + \frac{A^2\bar{v}}{16DL_1^3}\frac{(9-r^2)}{(r^3-r^2)}\right\} ln3 \tag{20}$$

To maximize the power in eq. (20), $r$ is set to be controllable such that $L_1$ and $\bar{v}$ are fixed. The maximum condition $\left(\frac{\partial p^*}{\partial r}\right)_{r=r_m} = 0$, where $p^*$ is the dimensionless power output given that all the dimensions of the power output are set to *unity* [65].

$$p^* = \frac{(r-3)}{4(r^2-r)}ln3 - \frac{(9-r^2)}{16(r^3-r^2)}ln\left(\frac{1}{3}\right) \tag{21}$$

The maximization condition leads to the following equation,

$$-2r^8 + 9r^7 + 15r^6 - 67r^5 + 63r^4 - 18r^3 = 0$$

which has five real solutions $r_m^1 = -3$, $r_m^2 = 0$, $r_m^3 = 1$, $r_m^4 = \frac{11-\sqrt{73}}{4}$ and $r_m^5 = \frac{11+\sqrt{73}}{4}$. However, considering the condition in eq. (17), among the five solutions, the physical solution is seen to be $r_m^5$. Therefore, our value for efficiency $\eta^*$ at maximum power is:

$$\eta^* = 1 - \frac{3}{\frac{11+\sqrt{73}}{4}} \cong 0.3860 \tag{22}$$

The dimensionless power output $p^*$ as a function of the efficiency $\eta$ is derived as:

$$p^* = \frac{(1-\eta)(3\eta - \eta^2)}{2+\eta} \tag{23}$$

In comparison with other works, the derived efficiency in eq. (22) proves that there is no generalized efficiency for the power of any Quantum system, but its values depend on the form of the potential [22,48,49].

From our derived dimensionless power output $p^*$ in eq (23), we plot its characteristics curve as a function of efficiency $\eta$ as shown in *figure 2*. Our result shows that there are two different efficiencies $\eta_1$ and $\eta_2$ at the maximum value of $p^*$. These efficiencies $\eta_1$ and $\eta_2$ represents regions where $\eta < \eta^*$ and $\eta > \eta^*$ respectively, we see that the lower efficiency $\eta_1$ only increases with a continual increase in the dimensionless power output $p^*$ indicating that it is not an optimal value for the heat engine. However, when our engine is operated in the later region where efficiency $\eta > \eta^*$, the dimensionless power output $p^*$ decreases as the efficiency increases, this region is often referred to as the ideal optimal value of an engine's efficiency:

$$\eta^* \leq \eta < 1$$

From the optimal region of our graph, it is obvious that the value of ratio $r$ is of interest to our engine when:

$$r^* \leq r \equiv r_m^4$$

Therefore, for the engine to operate at an optimal level, the condition $L_3 = r^* L_1$ must be satisfied.

Thus, the values of $\eta^*$ and $r^*$ are particularly important in determining the allowable value of the lower bound of the optimal efficiency and the structure of the engine, respectively.

## IV.  OUR RESULTS

The Morse oscillator potential [MO] is known to harmoniously transform into harmonic oscillator potential [HO] under several conditions, one of which is the divergence of its dissociation energy $D_0 \to \infty$ when the [MO] approaches the [HO]-limit. Inserting this condition in our results in equ. (20) and (23), give:

$$p = \frac{Av}{4L_1^2} \cdot \frac{(r-3)}{(r^2-r)} \ln 3 \tag{24}$$

$$p^* = \frac{(1-\eta)\eta}{2+\eta} \tag{25}$$

The results in equations (24) and (25) are the exact solution of our quantum heat engine's efficiency, power and dimensionless power respectively given that our working substance is a harmonic oscillator [HO] [49]. Nonetheless, the ability of the Morse oscillator to limit to a Harmonic oscillator does not mean that they exhibit similar characteristics when they are applied as a working substance of a Quantum system, rather *Figure 2* illustrates our findings, indicating that the Morse Oscillator [MO] exhibits superior power generation, reaching a maximum power of 0.26 with an efficiency of 40%. In contrast, the Harmonic Oscillator [HO] that yields a maximum power output of 0.10. Notably, the Morse Oscillator demonstrates heightened efficiency in the optimal region $\eta_2$, where $r^* \leq r$, making it a viable choice for Quantum systems. The efficiency of the Morse potential in this region underscores its effectiveness in harnessing energy for quantum processes. Additionally, *Figure 3* highlights that despite the Harmonic Oscillator potential showcasing higher dimensionless power peaks concerning the length ratio, the Morse oscillator potential proves more advantageous for Quantum systems anticipated to be nanoscopic. This is evident in the smaller size of the ratio $L_3 = r^* L_1$ within the engine's optimal region.

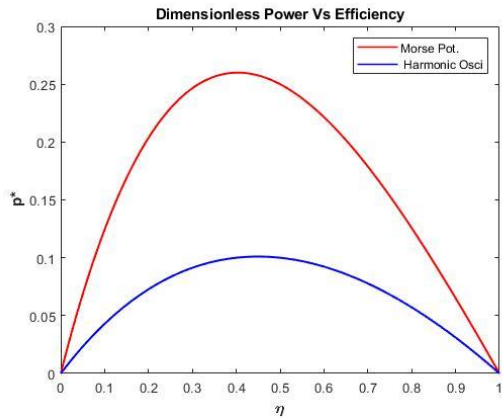
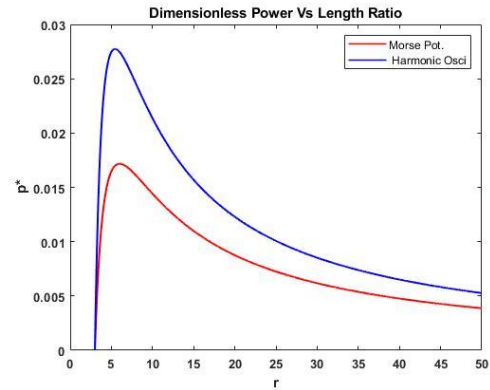

*Figure 2: The dimensionless power $p^*$ vs efficiency $\eta$*

*Figure 3: The dimensionless power $p^*$ vs width Ratio r.*

## V. CONCLUSION

In this study, we examined the optimization region for a quantum heat engine utilizing the Morse Oscillator potential as its working substance. Our findings reveal that the energy efficiency of Quantum heat engines is universally constant, independent of any parameters, while the efficiency of its power is contingent upon the potential's form and is constrained by the Carnot engine's value. Notably, this power efficiency is analogous to the well-established efficiency observed in classical thermodynamics. Our investigation, specifically focused on the Harmonic oscillator, yielded exact results, aligning with prior research. This underscores the uniqueness of the Morse Oscillator potential, urging further exploration as a promising working substance for Quantum heat engines.

## CONFLICT OF INTEREST

The authors declare no conflict of interest related to this work.

## DATA AVAILABILITY STATEMENT

The data that supports the findings of this paper is available from the corresponding author upon reasonable request.

The performance analysis of a Quantum-Mechanical Carnot-like Engine using Diatomic Molecules